%% file: ms.tex
\documentclass[conference,usenames,table,xcdraw]{IEEEtran}
\usepackage{multirow}
\usepackage{framed}
\usepackage{graphicx}
\usepackage{subfigure}
\usepackage{pifont}
\usepackage{xspace}
\usepackage{multirow}
\usepackage{comment}
\usepackage{xcolor}
\usepackage{hhline}
\usepackage{url}
\usepackage{booktabs} 
\usepackage{balance}
\usepackage{cite}
\include{macro}

\newcommand{\ie}{\textit{i.e.,}\xspace}
\newcommand{\eg}{\textit{e.g.,}\xspace}

\newcommand{\etal}{\textit{et al.}\xspace}

\def\hlinewd#1{%
\noalign{\ifnum0=`}\fi\hrule \@height #1 %
\futurelet\reserved@a\@xhline} 

\begin{document}

\title{Assessing Test Case Prioritization \\on Real Faults and Mutants}

\author{\IEEEauthorblockN{Qi Luo, Kevin Moran, Denys Poshyvanyk}
\IEEEauthorblockA{Department of Computer Science\\
College of William \& Mary\\
Williamsburg, VA\\
Email: \{qluo,kpmoran,denys\}@cs.wm.edu}
\and
\IEEEauthorblockN{Massimiliano Di Penta}
\IEEEauthorblockA{Department of Engineering\\
University of Sannio\\
Benevento, Italy\\
Email: dipenta@unisannio.it}}

\maketitle

\begin{abstract}
	Test Case Prioritization (TCP) is an important component of regression testing, allowing for earlier detection of faults or helping to reduce testing time and cost. While several TCP approaches exist in the research literature, a growing number of studies have evaluated them against synthetic software defects, called \textit{mutants}. Hence, it is currently unclear to what extent TCP performance on {\em mutants} would be representative of the performance achieved on \textit{real faults}. To answer this fundamental question, we conduct the first empirical study comparing the performance of TCP techniques applied to both real-world and mutation faults. The context of our study includes eight well-studied TCP approaches, 35k+ mutation faults, and 357 real-world faults from five Java systems in the Defects4J dataset.  Our results indicate that the relative performance of the studied TCP techniques on mutants may not strongly correlate with performance on real faults, depending upon attributes of the subject programs.  This suggests that, in certain contexts, the best performing technique on a set of mutants may not be the best technique in practice when applied to real faults. We also illustrate that these correlations vary for mutants generated by different operators depending on whether chosen operators reflect typical faults of a subject program.  This highlights the importance, particularly for TCP, of developing mutation operators tailored for specific program domains.
\end{abstract}

\IEEEpeerreviewmaketitle

\input{intro}

\input{background}

\input{study}

\input{results}

\input{threats}

\input{lessons}

\input{conclusion}

\balance
\bibliographystyle{abbrv}
\bibliography{ms}

\end{document}

%% file: intro.tex
\section{Introduction}
 
\textit{Regression Testing} is defined as the process of running a collection of compact tests, aimed at testing discrete functionality that underlies a software program, when that program changes in an evolutionary context.  This type of testing allows for the discovery of software \textit{regressions}, faults that cause an existing feature to cease functioning as expected.  
Regression test suites tend to be large for complex projects and are often run every time code is checked into a repository, leading to longer than desired testing times in practice. For example, Google has reported that, across its code bases, there are more than twenty code changes per minute, with 50\% of the files changing per month, leading to long testing times \cite{GoogleReport,Dini:ISSRE16}. To limit the number of test cases to execute,  \textit{Test Case Prioritization} (TCP) has been developed. TCP aims to prioritize test cases in a test suite to detect regressions more quickly or reduce testing time.

A large body of research has been dedicated to designing and evaluating  TCP techniques \cite{Arafeen:ICST13,Busjaeger:FSE16,Catal:2013,Czerwonka:ICST11,Do:08,Do:05,Kasurinen:2010, Liang:ICSE18,Miranda:ICSE18,Qu:ISSTA08,Wang:SPLC2014,Wang:FSE17,Yoo:STVR12,You:2011}. Such evaluations have typically been conducted by comparing the Average Percentage of Faults Detected (APFD) metric, or cost cognizant APFDc metric \cite{Elbaum:ICSE01,Epitropakis:ISSTA15,Jiang:09,Rothermel:99}, for different techniques.
 
 While in principle the evaluation of TCP techniques requires the availability of real program faults,
 very often the lack of real fault data has encouraged researchers to use artificial faults, called \textit{mutants}, each comprised of a simple syntactic change to the source code \cite{Do:06,Hao:TOSEM14,Henard:ICSE16,Lu:ICSE16,zhang2013bridging}.  The underlying assumption of such evaluations is that there is a strong correlation between prioritized sets of test cases that kill high numbers of mutants and sets that detect a high number of real faults.  This assumption raises key questions: \emph{How well do TCP techniques perform on real faults?}; \emph{Is the performance of TCP techniques on mutants representative of their performance on real faults?}; \emph{What properties of mutants affect the representativeness of this performance?} 
 
 Previous studies have examined the relationship between real faults and mutants in order to understand the applicability of mutants in software testing. In particular, these studies have investigated: (i) whether mutants are as difficult to detect as real faults \cite{AndrewsBL05,AndrewsBLN06}; (ii) whether mutant detection correlates with real fault detection \cite{DaranT96,Just:FSE14,Papadakis:ICSE'18}; (iii) whether mutants can be used to guide test case generation \cite{Shamshiri:ASE15}; and (iv) whether tokens contained in patches for real-world faults can be expressed in terms of mutants \cite{Gopinath:ISSRE14}.
 
Despite the studies outlined above, no previous study has investigated whether mutants are representative of real faults in the context of evaluating TCP approaches. Indeed, such evaluations aim to measure the \textit{rate} at which large sets of mutants are detected by \textit{prioritized sets of test cases} according to APFD(c), fundamentally differing from the experimental parameters of the studies outlined above. For example, Just {\em et al.} \cite{Just:FSE14} focused on the relationship between real faults and mutants measured by the ability of fault detection for a \emph{whole test suite}, which may not imply a similar relationship in terms of APFD(c) values. Furthermore, TCP approaches have not previously been extensively evaluated in terms of their capabilities for detecting real-world faults, indicating that the practical performance of these techniques is largely unknown.

To address this  research gap, we perform an extensive empirical study to understand the effectiveness of TCP techniques when evaluated in terms of real faults, and examine whether mutants are representative of real faults when evaluating TCP performance. We implemented eight well-studied TCP techniques and applied them on (i) a dataset of real faults, Defects4J \cite{Just:ISSTA14}, containing 357 real faults from five large Java programs, and (ii) over 35k+ mutants seeded using Pit \cite{PIT}. 

The results of our study show that  for the subject programs studied, mutation-based performance of TCP techniques, as measured by APFD(c), tends to \textit{overestimate} performance  \revision{by $\approx 20\%$ on average} when compared to performance on real faults \textit{unless trivial and subsumed mutants are removed}. When trivial and subsumed mutants are controlled for, performance measured according to the resulting mutants tends to slightly \textit{underestimate} performance \revision{by $ \approx3\%$ on average} compared to real faults.  Furthermore, our results indicate that the performance of TCP techniques \textit{relative to one another} on mutants, may not correlate to relative performance on real faults. However, the above findings tend to vary depending on the subject program. Additionally, we find that, as a whole, \revision{static TCP techniques tend to perform slightly better than dynamic techniques on real faults, but differences are not statistically significant}. Finally, when examining the fault sets in terms of mutant coupling and operator type, we found that the representativeness of mutants (compared to real faults) varies \textit{within} programs. This suggests that different mutant combinations could be derived to more closely resemble likely real faults in a program.   

To the best of our knowledge, this is the first comprehensive empirical study that evaluates the performance of eight well-studied TCP techniques on a large set of \textit{real faults} and compares the results to mutation-based performance in order to determine the representativeness of mutants in this context. \revision{The results of this study, and the code utilized are available in an online appendix to facilitate reproducibility \cite{appendix}.}

%% file: background.tex
\section{Background \& Related Work}

\label{sec:background}
In this section we formally define the TCP problem, introduce the studied TCP techniques, and discuss related work.

\subsection{TCP Problem Formulation}
\label{subsec:tcp}
TCP is formally described by Rothermel \etal \cite{Rothermel:TSE01} as finding a prioritized set of test cases $T'\in P(T)$, such that $\forall'', T''\in P(T)\wedge T''\neq T'\Rightarrow f(T')\ge f(T'')$, where $P(T)$ refers to the set of permutations of a given test suite $T$, and $f$ refers to a function from $P(T)$ to real numbers. While there are many types of existing TCP techniques \cite{do06nov,Elbaum:00,Hao:TOSEM14,Kavitha:10,Srivastava:02,Zhang:ICSM11}, one common dichotomous classification, \textit{static} \cite{Islam:CSMR12,Ledru:ASE12,Zhang:ICSM09} and \textit{dynamic} techniques \cite{Do:04,Elbaum:2003,Elbaum:SQI04,Korel:05,Nguyen:ICWS11,Smith:SAC09,Walcott:06,zhang2013bridging}, relates to the type of information used to perform the prioritization.  Static approaches utilize information extracted from source and test code and dynamic techniques rely on information collected at runtime (\eg coverage information per test case) to prioritize test cases \cite{Luo:FSE16}. Additional classifications exist, such as the distinction between \textit{white-box} and \textit{black-box} techniques \cite{Henard:ICSE16}.  The techniques that require source code of subject programs (or information extracted from source code) are typically classified as \textit{white-box} techniques \cite{Kapfhammer:2007}.  Conversely, those that only require program input or output information are classified as \textit{black-box} techniques. Approaches that only require test-code have been classified as black-box \cite{Henard:ICSE16,Luo:FSE16}, however, in this paper we more accurately refer to these techniques as \textit{grey-box} since they require access to test code with references to the underlying program.  There are also other approaches \cite{Jiang:compsac13,Saha:ICSE15,Varun:ICJST10} that use other types of information to perform the prioritization, such as code-changes and requirements, that do not fall neatly into these categories.
 
\subsection{Studied TCP Techniques}
\label{subsec:techniques}

In the context of our empirical investigation, we selected eight well-studied white and grey-box TCP techniques \revision{as they provide a effective means of comparing the performance of widely disseminated techniques on real faults to their reported performance on mutants in prior work. Furthermore, the original papers that introduce our studied approaches have been collectively cited over $1.4k$ times, indicating their importance within the research community.} Together we consider four dynamic white-box techniques that utilize run-time code coverage for prioritization~\cite{Jiang:09,Li:07,Rothermel:99}, two static gray-box techniques that operate only on test code~\cite{Ledru:ASE12,Thomas:EMSE14}, and two static white-box approaches that use call-graph information~\cite{Zhang:ICSM09}. In the remainder of this subsection we give an overview of these eight techniques (where the \textit{Greedy} and \textit{Call-Graph-based} techniques have two variants).  Implementation details for these techniques are given in Section~\ref{subsection:setup}.

\textit{1-2) \textbf{Greedy Techniques (Total \& Additional)}} The traditional greedy TCP techniques use two strategies, \textit{total} and \textit{additional}, to prioritize test cases using coverage information \cite{Rothermel:99}. The total strategy always prioritizes test cases with highest total coverage. Conversely, the additional strategy prioritizes tests that cover the most \textit{additional} code compared to the prioritized set.

\textit{3) \textbf{Adaptive Random Testing}} The Adaptive Random Testing (ART)-based TCP technique was proposed by Jiang \etal \cite{Jiang:09}. ART approaches are used to spread test inputs evenly over the entire input domain of a program in order to detect failures more quickly as compared to random testing \cite{Chen:Springer04}.  Initially, this technique randomly generates a candidate set of test cases and randomly selects a test case. Then, a new candidate set is iteratively generated in a random fashion and the test case that is farthest away from the already-prioritized set, in terms of coverage-based Jaccard distance, is added. In our study, we choose the \textit{minimum} distance to measure the distance between each test case and the set of prioritized test cases, since it has been shown to be the most effective \cite{Jiang:09}.

\textit{4) \textbf{\revision{Genetic Algorithm-based Technique}}} Li \etal proposed a set of approaches that prioritize test cases using search-based techniques, which are able to avoid producing sub-optimal results that find only local solutions within the test input space \cite{Li:07}. They used two meta-heuristics, namely hill climbing and Genetic Algorithms (GAs), and used coverage information as a prioritization objective. Their experimental results show that GAs perform better than hill climbing, thus, we utilize the GA-based approach in this study.

\textit{5-6) \textbf{Call-Graph-based Techniques (Total \& Additional)}}
Zhang \etal proposed an approach to prioritize test cases based on call-graph information instead of coverage \cite{Zhang:ICSM09}. The approach builds a call-graph for each test case, and uses the information to measure the testing abilities for all test cases. The test cases covering more methods are favored in the prioritization scheme. Similar to greedy techniques, there are two variations: \textit{total} and \textit{additional}.

\textit{7) \textbf{String-based Technique}} Ledru \etal proposed a TCP approach that uses textual information extracted from test cases \cite{Ledru:ASE12}. The underlying idea is that a set of textually diverse test cases may have a better chance at uncovering more faults than a textually similar set. This technique treats each test case as a string, and introduces four types of string distances to estimate the difference between a pair of test cases, prioritizing test cases that are furthest from the prioritized set in a pairwise manner. Experimental results show that Manhattan distance performs best in terms of fault detection \cite{Ledru:ASE12}, and thus, we use this setting in our study. 

\textit{8) \textbf{Topic-Model-based Technique}} This technique aims to utilize the textual information in test cases, such as identifiers and comments to build topic models for prioritization \cite{Thomas:EMSE14}. The technique relies on topic models to approximate the abstract functionality of each test case, and estimates the distances between topics of test cases to favor those that are able to test different high-level functionalities of the program. Similar to the string-based TCP technique, the topic-model-based technique uses Manhattan distance to measure the distance between a single test case and the prioritized set. The test furthest away from the prioritized set is chosen during each iteration to maximize diversity. In our study, we implement this approach as proposed by Thomas \etal~\cite{Thomas:EMSE14, Chen:13}. 

\vspace{-0.1cm}
\subsection{Threats to the Validity of Mutation-Based TCP Performance Evaluations}
\vspace{-0.1cm}
\label{subsec:tcp-threats}

There is a large body of work that has addressed the problem of Test Case Prioritization  \cite{Elbaum:TSE02,Jiang:compsac13,Mei:TSE12,Saha:ICSE15,Xu:ICST10}. A common link between the evaluations of various techniques is the utilization of \textit{fault-detection rates}, typically in terms of the APFD(c) metrics \cite{Henard:ICSE16,Lu:ICSE16,Tonella:ICSM06}. However, due to the fact that finding and extracting real-world faults from software is an intellectually intensive and laborious task, real faults are rarely used when evaluating testing related research \cite{Just:ISSTA14,Just:FSE14}, including existing work on TCP.  Instead, \textit{mutation analysis} can be utilized.
During mutation analysis small, automatically-generated syntactic faults are seeded throughout subject programs according to a set of clearly-specified \textit{mutation operators}. Then APFD(c) values are calculated according to the number of mutants that are detected (\ie killed) by prioritized test cases.  
To make results of such an evaluation generalizable to realistic settings,  
APFD(c) fault-detection rates for mutants should correlate with detection rates of \textit{real faults}.  Unfortunately, in the context of the typical methodology used to evaluate TCP techniques, the relation between performance on mutants and real faults is not well understood.  This gives rise to the potential for threats to the validity of TCP evaluations relating to (i) the overall performance of TCP techniques on mutation vs. real faults (\textit{is performance over or under-estimated?}); (ii) the relative performance or performance correlation of real vs. mutation faults (\textit{does a mutation based-analysis properly illustrate the most effective technique on real faults?}); and (iii) the impact that different properties of mutants and programs have on mutation-based performance of TCP.

\vspace{-0.1cm}
\subsection{Studies Examining the Relationship Between Mutants and Real Faults}
\vspace{-0.1cm}
\label{subsec:real-fault-studies}

While our study is one of the first to examine the representativeness of mutation faults in terms of real faults as it pertains to the domain of TCP, we are not the first to investigate this relationship in a general sense. Daran and Th\'evenod-Fosse \cite{DaranT96} performed the first empirical comparison between  mutants and real faults, finding that the set of errors and failures they produced with a given test suite are quite similar. Andrews \etal \cite{AndrewsBL05,AndrewsBLN06} compared the fault detection capability of test-suites on mutants, real-faults, and hand-seeded faults, reaching two conclusions.  First, mutants (if carefully selected) can provide a good indication of a test suite's ability to detect real faults. Second, the use of \textit{hand-seeded} faults can produce an underestimation of a test suite's fault detection capability. Gopinath \etal conducted an empirical study that explored the characteristics of a large set of changes and bug-fixes and how these related to mutants \cite{Gopinath:ISSRE14}. 

Just \etal studied whether a test suite's ability to detect mutants is coupled with its ability to detect real faults, controlling for code-coverage \cite{Just:FSE14}. Their results indicate that mutant detection correlates more closely with real fault detection than with code coverage. Additionally, their study also provided suggestions regarding how mutant taxonomies can be improved to make them more representative of real faults through examination of how mutants are coupled to real faults. Just \etal introduced a valuable dataset of  357 real faults across five Java programs in an artifact called Defects4~\cite{Just:ISSTA14}, which we utilize in this paper. Shamshiri \etal conducted an empirical study of automatic test generation techniques to investigate their ability to detect real faults in the Defects4J~\cite{Shamshiri:ASE15}.  Finally, Chekham \etal conducted a study examining how mutation, statement and branch coverage correlate to fault revelation \cite{Chekam:ICSE'17}. They found that \textit{strong} mutation testing has the highest fault revelation capability and fault revelation only tends to significantly increase once high coverage levels are attained. 

While the aforementioned research has investigated several aspects of the relationship between real-faults and mutants, there is very little prior work examining this relationship in the context of typical TCP evaluation methodologies. Thus, it is unclear the extent to which many of the results from these prior studies hold in the context of TCP, as the experimental settings in such cases (and hence in our study) \textit{fundamentally differ} from past work.  These differences manifest in the typical methodology used to assess the effectiveness of TCP techniques, which involves seeding mutants into a \textit{single} version of a program and calculating fault-detection \textit{rates} according to the APFD(c) metrics.

\revision{In order to more clearly illustrate the need for this study, it is useful to consider how our experimental setup differs from prior work such as Just et. al's~\cite{Just:FSE14}.  The most analogous study to ours that Just et. al conduct is that which measures the correlation between a test suite's ability to detect real faults and the test suite's mutation score, \textit{without considering the impact of test case ordering within a suite}.  In contrast, our study is, in essence, aimed at investigating how test case ordering (and prioritization) can impact the representativeness of mutants in terms of fault detection \textit{rates}. This nuance has important implications for TCP because the \textit{distribution} of faults across a program can impact the rate at which a test suite detects faults depending on test case ordering. For example, mutants could be distributed throughout a program in a more uniform manner than real faults, causing differences in the effectiveness of the prioritization schemes of different TCP techniques. Previous studies that only examine test suites in their entirety were not capable of exploring this phenomenon.} 

	It should be noted that while this paper was under review, Patterson \etal \cite{Paterson:AST'18} published a study examining the effect that single and multiple real faults and mutants have on the effectiveness of four TCP techniques, and concluded that mutants may not be an effective surrogate for real faults. Our study complements this work as we consider a larger number of Defects4J faults, additional TCP techniques, and the effect of removing subsumed and trivial mutants.

%% file: study.tex
\section{Empirical Study}
\label{sec:study}

The {\em goal} of this study is to analyze the extent to which mutation analysis can support the evaluation of Test Case Prioritization, as opposed to using data from real faults. The study {\em context} consists of data from five Java open source projects (Defect4J~\cite{Just:ISSTA14,Just:FSE14}), mutants generated by PIT \cite{PIT}, and the eight TCP techniques described in Section \ref{sec:background}.
\input{tables/subs}

\subsection{Research Questions (RQs):}
\label{subsec:rqs}

The study aims at answering the following three \textbf{RQs}:

\begin{itemize}
\item \textbf{RQ$_1$:} How \textit{effective} are TCP techniques when applied to detecting real faults?
\item \textbf{RQ$_2$:} Is the performance of TCP techniques on mutants \textit{representative} of performance on real faults? 
\item \textbf{RQ$_3$:} How do the \textit{properties} of real faults and mutants affect the performance of TCP techniques? 
\end{itemize}

\subsection{Study Context}
\label{subsec:context}

In order to properly evaluate the performance of TCP approaches when applied to detecting real-world faults, our study requires a well understood set of verified, real program faults preferably containing coupling information between real faults and mutants.  To satisfy this criteria, we utilize the Defects4J \cite{Just:ISSTA14} dataset, which contains 357 real faults extracted from five Java subject programs, listed in Table \ref{table:sub}, and has been utilized in past studies \cite{Just:FSE14,Shamshiri:ASE15}. Defects4J isolates the real faults from the version control and bug tracking system of each subject program. For each isolated fault there exists a faulty program version and a corresponding fixed version. Table \ref{table:sub} shows the distribution of isolated real faults across the five subject programs, with Closure Compiler and Commons Maths representing the largest numbers of faults.

For each real fault (\ie faulty version), Defects4J provides a test suite including at least one test case that is able to trigger the fault but pass successfully in the corresponding fixed version. Additionally, it provides the code locations (\ie method and class names) that were modified to fix the fault. Test cases were extracted at test-method granularity rather than the test-class granularity, as TCP techniques have been shown to perform best under such experimental settings \cite{Luo:FSE16}.

The \textit{primary goal} of this study is to determine how well mutation-based measures of TCP effectiveness reflect the performance of these techniques on real faults. More generally, we aim to answer the following question: ``If one prioritizes test cases using mutants, would this prioritized set likely be as effective on real faults?"  In order to explore this question we seeded mutants using the PIT mutation tool \cite{PIT} with all built-in operators enabled. During this seeding process we excluded mutants that cannot be killed (\ie triggered the test case to fail), by any test case in the existing JUnit suites for two reasons: i) to mitigate a potential threat to validity from equivalent mutants, ii) they do not affect our studied metrics according to the definitions of APFD(c) as defined in Section \ref{subsec:methodology}. The number of detected mutants and the total number of seeded mutants are shown in columns 3 and 5 of Table \ref{table:sub} respectively (see online appendix for further information~\cite{appendix}). 

	To perform mutant seeding that allows for comparison between real faults and mutants, for each (real) \textit{faulty} version of a program in Defects4J, we create one mutated program instance by seeding a randomly selected mutant into the \textit{latest} corresponding program version, and then repeat this process 100 times (\eg for Closure: 133 versions with real faults $\times$ 1 mutant $\times$ 100 instances = 13,300 total mutants). For instance, taking the JFreeChart program as an example, one mutant was randomly selected from the set of 32,790 mutants able to be detected by at least one test case, until a set of 26 mutant versions of JFreeChart were accumulated (matching the number of real faulty versions). This procedure is then repeated 100 times. This results in 100 groups of 26 mutants, or 2,600 mutants being evaluated for JFreeChart in total. We repeat the seeding process 100 times due to the fact that the selected mutant is a random variable, and we aim to provide a reliable statistical analysis and the best possible approximation for TCP evaluations from prior work.   In initial experiments, excluded due to space limitations (but included in our online appendix~\cite{appendix}), we computed the APFD(c) values using five randomly seeded mutants per instance (instead of one), following the settings of previous work \cite{Hao:TOSEM14,Lu:ICSE16,Luo:FSE16,Mei:TSE12,Zhang:ASE13}. The results for this analysis generally agree with the presented results, and thus we do not expect that the number of mutants per instance will dramatically impact findings. 

	The intention for choosing these experimental settings is to evaluate whether past mutant-based \textit{methodologies} measuring TCP efficacy hold for real faults. Thus, we seeded mutants in accordance with past studies \cite{Luo:FSE16,Mei:TSE12,Saha:ICSE15,Zhang:ASE13,zhang2013bridging}, applying mutation analysis to the \textit{latest} version of each subject program. This means that test suites from each faulty program version in Defects4J are prioritized for both each faulty version and the latest version (with mutants injected). As Section \ref{sec:threats} describes, this makes for a reasonable comparison due to the fact that past work measuring the impact of software evolution on TCP efficacy has shown that mutation-based performance remains consistent when applying test cases from an earlier program version to both the earlier and later versions~\cite{Lu:ICSE16}.

	Furthermore, two recent works outline the potential impact of trivial/subsumed mutants for mutation-based analysis \cite{Henard:ICSE16,Papadakis:ISSTA16}. Thus, we investigate our RQs both with and without trivial and subsumed mutants removed from the set of seeded mutants.  We follow the methodology defined  in prior work \cite{Papadakis:ISSTA16}, which is the best approximation for the removal of subsuming mutants, as this has been proven an undecidable problem. The number of subsuming mutants is shown in Table~\ref{table:sub}. Thus we will discuss results in terms of two different mutant \textit{sets}: the \textit{full mutant set}, and the \textit{subsuming mutant set}. 

\vspace{-0.15cm}
\subsection{Methodology}
\label{subsec:methodology}

\input{tables/strategies}

\subsubsection{\textbf{RQ$_{1}$: TCP Effectiveness on Real Faults}}

The \textit{goal} of this research question is to investigate the performance of TCP techniques when they are applied to detect real faults. We first ran these eight TCP techniques on 357 Defects4J program versions containing real faults to obtain ranked lists of test cases for each faulty version. The tests are run at the \textit{test-method} level, since past work has shown method-level yields more effective TCP results \cite{Lu:ICSE16, Luo:FSE16}.  Then, to measure the effectiveness in terms of fault detection for each studied technique,
we calculated two well-accepted metrics, the Average Percentage of Faults Detected (APFD) \cite{Elbaum:TSE02,Rothermel:99} and its cost cognizant counterpart APFDc~\cite{Elbaum:ICSE01,Epitropakis:ISSTA15}. Formally, APFD is defined as follows: Let $T$ be a test suite and $T'$ is a permutation of $T$, the APFD value for $T'$ is  given by
\begin{equation}\label{Equ:APFD}
    APFD=1- \frac{\sum_{i=1}^{m}TF_i}{n \times m}+\frac{1}{2n}
\end{equation}
where $n$ is the number of test cases in $T$, $m$ is the number of faults, and $TF_{i}$ is the position of the first test case in $T'$ that detects fault $i$. Intuitively, the higher the APFD value, the higher the rate of fault detection by the prioritized test cases. In order to derive a more holistic understanding of the relationship between TCP performance on real faults and mutants we also consider APFDc.  This metric takes both execution cost and fault severity into account. Since there is no clearly-defined nor widely-used method for estimating fault severity, we consider severity to be the same for all faults. Therefore, in the context of this study APFDc reduces to the following formal definition:
\begin{equation}\label{Equ:APFDC}
	APFDc=\frac{\sum_{i=1}^{m}\left(\sum_{j=TF_i}^{n}t_j - \frac{1}{2}t_{TF_i}\right)}{\sum_{j=1}^{n}t_j \times m}
\end{equation}  
where $t_j$ is the execution cost for the test case ranked at position $j$ in the ranked test suite. Intuitively, APFDc, as defined above, will be higher for prioritized test suites that both find faults faster and require less execution time.  Since we study five subjects, each with a different number of real faults, we computed the average APFD(c) values of the different versions across all five systems to understand the effectiveness of each studied approach. Additionally, to statistically analyze the differences between TCP techniques in terms of APFD and APFDc values, we perform an Analysis of Variance (ANOVA) and a Tukey Honest Significant Difference (HSD) test \cite{tukey} on the average APFD(c) values across the five subjects. The ANOVA analysis is used to test whether there are statistically significant differences between the performance of TCP techniques when applied to real faults versus when applied to mutants. The Tukey HSD test classifies the TCP techniques into different groups based on their performance in terms of APFD(c) values. For both statistical procedures we consider a significance level $\alpha=0.05$.

\subsubsection{\textbf{RQ$_{2}$: Representativeness of Mutants}}

	The \textit{goal} of RQ$_{2}$ is to understand whether mutants are representative of real faults in the evaluation of TCP techniques. Thus, we applied mutation analysis according to the description given in Section \ref{subsec:context}. As mentioned in Section \ref{subsec:context}, two sets of mutants are examined, the \textit{full mutant set} and the \textit{subsuming mutant set}. Then, we ran all studied TCP techniques on these sets of mutant versions and calculated the average APFD(c) values (see Equations \ref{Equ:APFD} and \ref{Equ:APFDC}) across all 100 mutant groups for all five subject programs, in order to examine the mutant-based performance of our studied TCP techniques.

	At this point, we are able to evaluate the effectiveness of TCP techniques in terms of both real fault and mutant detection according to APFD(c), which we refer to as the \textit{absolute performance}.  In addition to the absolute performance, we are also concerned with how different techniques perform relative to one another across different fault sets, which we refer to as the \textit{relative performance}.  If the ranking of TCP techniques from most to least effective is similar when measured across both mutants and real faults, the relative performance would be \textit{positively correlated}, otherwise it would be \textit{negatively correlated}. To calculate this, we utilize the Kendall rank correlation coefficient $\tau$ \cite{Kendall:1938} to measure the relationship. This correlation metric is commonly used to measure the ordinal association between two quantities \cite{Kendall:Wiki} and has been widely used in the area of software testing for such purpose \cite{Gligoric:ISSTA13,Papadakis:ICST14,Zhang:ASE13}. Consider the APFD values of real fault detection and mutant detection across all studied techniques as a set of pairs $(R, M)$, where $R$ is the APFD/APFDc values of real fault detection and $M$ is the APFD/APFDc values of mutant detection. Any pair of $(r_i, m_i)$ and $(r_j, m_j)$ (APFD values for TCP$_i$), where $i \neq j$, are concordant if $r_i > r_j$ and $m_i > r_j$ or if $r_i < r_j$ and $m_i < m_j$. They are discordant if $r_i > r_j$ and $m_i < m_j$ or if $r_i < r_j$ and $m_i > m_j$ \cite{Kendall:EOM}. The Kendall $\tau$ rank correlation coefficient is formally defined as the ratio of the number of concordant pairs less the number of discordant pairs and the total number of pairs. Thus, its value ranges from $-1.0$ to $1.0$. Results closer to $1.0$ indicate the observations of two variables have similar rank (\eg which in the context of this study translates to similar rates of fault discovery), whereas when it is closer to $-1.0$ when the observations of two variables have dissimilar ranks (\eg suggesting a negative correlation between  fault discovery rates). Following previous work \cite{Gligoric:ISSTA13}, we use the Kendall $\tau_{b}$ statistic as it accounts for ties and does not require a linear relationship.

\subsubsection{\textbf{RQ$_{3}$: Effects of Fault Properties}}
\label{subsubsec:fault-prop}

The \textit{goal} of this research question is to understand how different \textit{properties} of faults impact two phenomena in the context of TCP: (i) the performance of techniques, and (ii) the utility of mutants as a proxy for real faults (\ie performance correlation).   

The \textit{first} fault property we investigated is {\bf the level of coupling between real faults and mutants}.  In order to determine the level of coupling for real faults to mutation operators, we utilize Just \emph{et al.}'s previous work \cite{Just:FSE14}, which classified the 357 real faults from the Defects4J dataset into four coupling levels: (i) those coupled with mutants (denoted in the study using the keyword ``Couple"), (ii) those requiring stronger mutation operators (denoted as ``StrongerOP"), (iii) those requiring new mutation operators (denoted as ``newOP"), and (iv) and those not coupled with mutants (denoted as ``Limitation")  Formally speaking, a real fault (\ie a complex fault) is coupled with a set of mutants (\ie simple faults) if a test case that detects all the mutants also detects the real fault \cite{Just:FSE14}. We contacted the authors to obtain this classification scheme, which includes 262 real faults coupled to mutants, 25 real faults requiring stronger mutation operators, seven real faults requiring new mutation operators, and 63 real faults not coupled to mutants. In order to examine the impact that fault coupling has on \textit{performance}, we pruned our initial dataset from the previous two research questions and calculated APFD(c) values for each coupling level of real faults and for all mutants considered in RQ$_{2}$ for each subject program. In order to examine the \textit{correlation} between these APFD(c) values, we again utilize the Kendall $\tau_{b}$  coefficient.

The \textit{second} fault property we examined is the {\bf mutation operator type}. Intuitively, this investigation should help shed light on which mutation operators are more representative of real faults for our studied subject programs in the context of TCP performance.  We classified mutants based on their corresponding operators,  that is, the mutation faults that are generated by the same mutation operator are classified into the same group. We consider the 15 built-in operators in PIT \cite{PIT}, and for each subject program classified all mutant versions into groups according to these operators.  For each subject program and operator type, we then randomly sampled from these groups until we had a set of faults corresponding to the number of real faults existent in each subject program respectively. We then repeat this process 100 times. If there are not enough mutants to create 100 groups, we repeat this process until we exhaust the mutants. In the end, for each subject program, we derive 100 mutant groups for each type of operator (given enough mutants), with each group containing the same number of mutants (all of the same operator type) as real-faults for the subject. To understand the impact that different types of mutation operators have on the \textit{performance} of TCP techniques, we calculated the APFD(c) values based on the new groups of mutants, and then used the Kendall $\tau_b$ coefficient to understand the \textit{correlation} between the APFD(c) values calculated in terms of real faults and mutants. We also explored the effects of mutant locations, however, we found no significant trends. Thus, we forgo discussion of these results and point interested readers to our attached appendix.

\input{tables/overall-perf}

\subsection{Experiment Tools and Hardware}
\label{subsection:setup}

We reimplemented all studied TCP techniques in accordance with the technical descriptions in their corresponding papers (see Sec. \ref{sec:background}). Three of the authors, and an external expert on TCP, carefully reviewed the code, ensuring the reimplementation is reliable. To collect coverage information, we used the ASM bytecode manipulation and analysis toolset \cite{ASM}. In our empirical study, we chose to use statement-level coverage information, as this allows for optimal performance of TCP techniques \cite{Luo:FSE16}. Furthermore, we utilize JDT \cite{JDT} to extract textual information for each test method, which is used by string-based and topic-based approaches. Specifically for the topic-based approach, we use Mallet \cite{Mallet} to build a latent Dirichlet allocation (LDA) topic model \cite{LDA} for each test case, after pre-processing the textual information (\eg splitting, removing stop words and stemming). Following previous research \cite{Luo:FSE16}, we use WALA \cite{WALA} to build RTA call graphs \cite{Grove:OOPSLA97} for each test method and traverse each call graph to obtain its static coverage to implement the $TCP_{cg}$ techniques.

The experiments were carried out on eight servers with 16, 3.3 GHz Intel Xeon E5-4627 CPUs and 512 GB RAM, and one server with eight Intel X5672 CPUs and 192 GB RAM.

%% file: tables/subs.tex
\begin{table}
\small
\vspace{-0.6cm}
\setlength{\tabcolsep}{4.0pt}
\center\caption{\label{table:sub}\small The stats of the subject programs
}\vspace{-0.3cm}
\begin{tabular}{|l||c|c|c|c|c|}
\hline
Subject Programs&\#Real&{\#Detected}&{\#Subsuming}&\#All\\
\hline\hline
JFreeChart&26&32,790&1,796&102,629\\
Closure Compiler&133&82,572&9,731&111,826\\
Commons Maths&106&80,059&5,016&113,680\\
Joda-Time&27&24,555&3,066&34,147\\
Commons Lang&65&25,173&2,129&31,214\\
\hline\hline
Total&357&245,767&21,738&393,496\\\hline
\end{tabular}
\vspace{-0.5cm}
\end{table} 

%% file: tables/strategies.tex
\begin{table}
\vspace{-0.7cm}
\center
\setlength{\tabcolsep}{4.6pt}
\footnotesize
\caption{\label{table:tcp}Studied TCP Techniques.}
\vspace{-0.3cm}
\hspace{-0.05cm}
\begin{tabular}{|l||c|c|}
\hline
Type&Tag&Description\\
\hline
\hline
\multirow{4}{*}
{Static}&$TCP_{cg-tot}$&Call-graph-based (total strategy)\\
&$TCP_{cg-add}$&Call-graph-based  (additional strategy)\\
&$TCP_{str}$&The string-distance-based\\
&$TCP_{topic}$&Topic-model-based\\
\hline
\hline
\multirow{4}{*}
{Dynamic}&$TCP_{total}$&Greedy total  (statement-level)\\
&$TCP_{add}$&Greedy additional (statement-level)\\
&$TCP_{art}$&Adaptive random (statement-level)\\
&$TCP_{search}$&Search-based (statement-level)\\
\hline
\end{tabular}
\vspace{-0.45cm}
\end{table}

%% file: tables/overall-perf.tex
\begin{table*}[t]
\centering
\vspace{-1.2cm}
\footnotesize
\definecolor{mygray}{gray}{0.9}
\caption{Average APFD \& APFDc values for all eight TCP techniques, for real, full mutant and subsuming mutant fault sets, across all subject programs. Additionally, the grouping results for the Tukey HSD test are shown in capitalized letters (\eg AB). S.Mutants refers to subsuming mutants. \vspace{-0.2cm}}
\label{tab:overall-perf}
\resizebox{\linewidth}{!}{
\begin{tabular}{|c!{\vrule width 0.8pt}c|c!{\vrule width 0.8pt}c|c!{\vrule width 0.8pt}c|c!{\vrule width 0.8pt}c|c!{\vrule width 0.8pt}c|c!{\vrule width 0.8pt}c|c!{\vrule width 0.8pt}c|c!{\vrule width 0.8pt}c|c!{\vrule width 0.8pt}}
\hline
                        & \multicolumn{8}{c!{\vrule width 0.8pt}}{Static Techniques}        & \multicolumn{8}{c!{\vrule width 0.8pt}}{Dynamic Techniques} \\ \hline
Faults                        & \multicolumn{2}{c!{\vrule width 0.8pt}}{TCP$_{cg-tot}$}        & \multicolumn{2}{c!{\vrule width 0.8pt}}{TCP$_{cg-add}$}        & \multicolumn{2}{c!{\vrule width 0.8pt}}{TCP$_{str}$}           & \multicolumn{2}{c!{\vrule width 0.8pt}}{TCP$_{topic}$}         & \multicolumn{2}{c!{\vrule width 0.8pt}}{TCP$_{total}$}         & \multicolumn{2}{c!{\vrule width 0.8pt}}{TCP$_{add}$}           & \multicolumn{2}{c!{\vrule width 0.8pt}}{TCP$_{art}$}           & \multicolumn{2}{c!{\vrule width 0.8pt}}{TCP$_{search}$}        \\ \hline\hline
                              & APFD  & {\cellcolor{mygray}}APFDc & APFD  & {\cellcolor{mygray}}APFDc & APFD  & {\cellcolor{mygray}}APFDc & APFD  & {\cellcolor{mygray}}APFDc & APFD  & {\cellcolor{mygray}}APFDc & APFD  & {\cellcolor{mygray}}APFDc & APFD  & {\cellcolor{mygray}}APFDc & APFD  & {\cellcolor{mygray}}APFDc \\ \hlinewd{0.7pt}
                              & 0.594 & {\cellcolor{mygray}}0.480 & 0.597 & {\cellcolor{mygray}}0.591 & 0.696 & {\cellcolor{mygray}}0.594 & 0.7   & {\cellcolor{mygray}}0.635 & 0.61  & {\cellcolor{mygray}}0.419 & 0.583 & {\cellcolor{mygray}}0.454 & 0.657 & {\cellcolor{mygray}}0.677 & 0.6   & {\cellcolor{mygray}}0.556 \\ \hhline{l~|-|-|-|-|-|-|-|-|-|-|-|-|-|-|-|-|}
\multirow{-2}{*}{Real} & A     & {\cellcolor{mygray}}BC    & A     & {\cellcolor{mygray}}ABC   & A     & {\cellcolor{mygray}}ABC   & A     & {\cellcolor{mygray}}AB    & A     & {\cellcolor{mygray}}C     & A     & {\cellcolor{mygray}}C     & A     & {\cellcolor{mygray}}A     & A     & {\cellcolor{mygray}}ABC   \\ \hlinewd{0.7pt}
                              & 0.743 & {\cellcolor{mygray}}0.598     & 0.818 & {\cellcolor{mygray}}0.835 & 0.834 & {\cellcolor{mygray}}0.788 & 0.832 & {\cellcolor{mygray}}0.802 & 0.757 & {\cellcolor{mygray}}0.549 & 0.897 & {\cellcolor{mygray}}0.829 & 0.8   & {\cellcolor{mygray}}0.841 & 0.784 & {\cellcolor{mygray}}0.725 \\ \hhline{l~|-|-|-|-|-|-|-|-|-|-|-|-|-|-|-|-|}
\multirow{-2}{*}{Full Mutant} & B     & {\cellcolor{mygray}}BC    & AB    & {\cellcolor{mygray}}A     & AB    & {\cellcolor{mygray}}AB    & AB    & {\cellcolor{mygray}}A     & B     & {\cellcolor{mygray}}C     & A     & {\cellcolor{mygray}}A     & AB    & {\cellcolor{mygray}}A     & B     & {\cellcolor{mygray}}ABC   \\ \hlinewd{0.7pt}
                              & 0.561 & {\cellcolor{mygray}}0.407     & 0.612 & {\cellcolor{mygray}}0.639 & 0.620 & {\cellcolor{mygray}}0.572 & 0.612 & {\cellcolor{mygray}}0.570 & 0.534 & {\cellcolor{mygray}}0.305 & 0.664 & {\cellcolor{mygray}}0.565 & 0.622   & {\cellcolor{mygray}}0.671 & 0.578 & {\cellcolor{mygray}}0.508 \\ \hhline{l~|-|-|-|-|-|-|-|-|-|-|-|-|-|-|-|-|}
\multirow{-2}{*}{S.Mutant} & AB     & {\cellcolor{mygray}}BC   & AB    & {\cellcolor{mygray}}A     & AB    & {\cellcolor{mygray}}AB    & AB    & {\cellcolor{mygray}}AB     & B     & {\cellcolor{mygray}}C     & A     & {\cellcolor{mygray}}AB     & AB    & {\cellcolor{mygray}}A     & AB     & {\cellcolor{mygray}}ABC   \\ \hline
\end{tabular}
}
\vspace{-0.4cm}
\end{table*}

%% file: results.tex
\section{Results}
In this section, we describe the results of our empirical study.  Furthermore, we provide an online appendix with additional results~\cite{appendix}.

\subsection{RQ$_1$: TCP Effectiveness on Real Faults} 
\label{subsec:rq1}

The values of the APFD(c) metrics and the results of the Tukey HSD test for \textit{real} faults are reported at the top of Table \ref{tab:overall-perf}. From these experimental results we make the following observations. First, for real faults, all techniques tend to perform better when measured by APFD as compared to APFDc. This is not surprising, and it is due to the incorporation of execution cost. For some techniques, in particular $TCP_{total}$ and $TCP_{cg-tot}$, the differences between APFD and APFDc are comparatively larger. This observation is most likely due to the fact that these techniques always prioritize test cases with higher coverage first, leading to longer execution costs for the top test cases, in turn leading to lower APFDc values.

\input{tables/overall-correlation}

\revision{Second, the static TCP techniques perform slightly better overall as compared to dynamic TCP techniques for both APFD and APFDc metrics, but these differences are not statistically significant.  To determine whether static techniques outperformed dynamic ones to a statistically significant degree, we performed a Wilcoxon signed rank test across the raw APFD(c) values achieved by the collective sets of static and dynamic techniques across all subject systems. Results indicate that while static techniques do generally outperform dynamic techniques, the differences are not statistically significant, and the Cliff's delta effect sizes are negligible to small. We provide complete analysis results in our appendix \cite{appendix}}.

Third, the $TCP_{add}$ technique does not outperform the $TCP_{tot}$ strategy, contradicting findings from past studies where the $TCP_{add}$ has been shown to perform best overall \cite{Elbaum:TSE02,Luo:FSE16,Rothermel:TSE01}. Fourth, the results of the Tukey HSD test suggest that for APFD, the performance of the TCP programs does not vary in a statistically significant manner. However, for APFDc, we found statistically significant differences across techniques. It should be noted that the results of the statistical tests are derived from a smaller dataset as compared to the traditional method of using thousands of mutation faults, due to the number of faults included in Defects4J. 


\subsection{RQ$_2$: Representativeness of Mutants}
\vspace{-0.1cm}
\label{subsec:rq2}

\input{tables/coupling-correlation}

The values of the APFD(c) metrics for mutants across the different TCP techniques are also shown in Table \ref{tab:overall-perf}.
From this data, we make several notable observations.  First, the APFD and APFDc metrics calculated using the full mutant set  generally tend to \textit{overestimate} absolute performance compared to real faults \revision{by $\approx 20\%$ on average}. This finding is relevant, as it implies that mutation-based evaluations measuring the absolute performance of TCP techniques that do not control for subsumed and trivial mutants tend to overestimate \textit{real-world} absolute performance of these techniques. Conversely, there is a slight \textit{underestimation} \revision{($\approx 3\%$ on average)} for the APFD(c) values calculated using the subsuming mutant set when compared to real faults. Moreover, the studied TCP techniques perform differently across different fault sets, with the absolute performance of techniques on both the full mutant set and subsumed mutant set differing from absolute performance on real faults. This is a significant finding as it suggests that, according to results for our set of five subject programs, \textit{\textbf{a TCP approach that performs well according to mutation analysis may not exhibit the same performance on a set of real faults for the same program(s)}}. As we will discuss later, this points to the need for \textit{careful selection of mutants} when performing a mutation-based evaluation of TCP techniques in order to ensure that the results obtained also hold for \textit{real faults}. Additionally, removing subsumed mutants should prove beneficial in practice to avoid performance overestimation.

While the absolute performance in terms of APFD and APFDc may not be similar when comparing performance on mutants to performance on real faults, it is possible that performance is \textit{positively correlated} between the different fault sets.  That is, the performance of TCP techniques on real faults \textit{relative to each other} is consistent across different fault sets.  To measure this, we examine the Kendall $\tau_b$ correlation coefficient across the results from the two types of faults (shown in Table \ref{tab:overall-correlation}). Note that two rankings are considered as independent to one another when $\tau_b$ is closer to zero. Our results indicate a very weak positive correlation between mutants and real faults when examining APFD ($\tau_b$=0.143 for all-killed mutants and for $\tau_b$=-0.071 subsuming mutants) and a medium to strong positive correlation when considering APFDc  ($\tau_b$=0.571 for all-killed mutants and for $\tau_b$=0.643 subsuming mutants). Removing subsumed mutants does not impact the correlation results. This observation implies that, in general, a mutation-based TCP performance evaluation carried out in terms of APFDc will more strongly correlate to performance in terms of APFDc on real faults.  However, when relying on a mutation-analysis based APFD evaluation, \textit{as many previous studies do}, there is no guarantee that the results will correlate to similar levels of performance on real faults.  However, as we illustrate in the course of answering RQ$_3$, this correlation tends to vary across both the studied techniques and mutation operators.  \revision{Furthermore, the actual rankings of different techniques across the different fault sets can impact the significance of these results. For example, if the weak Kendall correlation for the APFD metric is merely due to slight shuffling of a set of consistently top-ranked techniques, this may indicate that the relative performance of techniques according to \textit{mutants} may be closer to real faults than implied by our correlation analysis.  However, this is not what we observe; for APFD, on average, the relative performance ordering according to real faults indicates that $TCP_{topic}$ performs best and $TCP_{add}$ performs worst. However, for both the full and subsuming mutant sets, $TCP_{add}$ performs best, representing a reversal in the ordering of the best and worst ranked techniques. We observe similar trends for APFDc.} 

\subsection{RQ$_3$: Effects of Fault Properties}
\label{subec:rq3}

\begin{figure}
\vspace{-0.4cm}
\flushleft
\includegraphics[width=0.9\columnwidth]{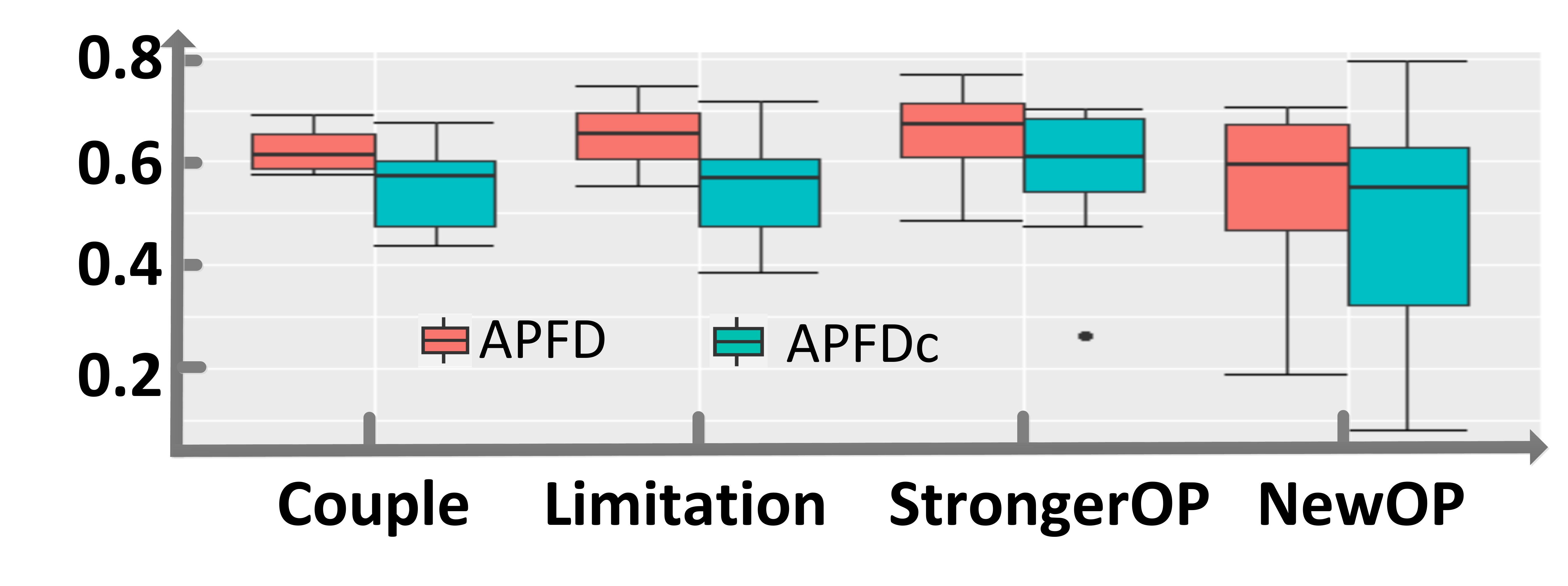}
\vspace{-0.3cm}
\caption{\small APFD(c) values for TCP techniques in terms of detecting different types of real faults.}
\label{fig:coupling-perf}
\vspace{-0.5cm}
\end{figure}

In this subsection we investigate the effect that different fault properties have on the performance of TCP techniques, and on the relationship between mutants and real faults.  As stated earlier, we discuss results for real faults in terms of different \textit{coupling levels}, and mutants in terms of \textit{operators}. In the context of \textbf{RQ$_{3}$}, we use the full mutant set for analysis instead of subsuming mutants, due to the limited number of subsuming mutants, particularly when grouping them based on mutation operators. However, interested readers can find the results for subsuming mutants in our attached appendix.

\subsubsection{\textbf{Effects of Coupling Between Mutants and Real Faults}}

To investigate the effect that fault coupling has on the performance of TCP across real-faults we consider four different fault types discussed in Section \ref{subsubsec:fault-prop}.  The \textit{performance} results for the APFD(c) metrics broken down by coupling level are illustrated in Figure \ref{fig:coupling-perf}. The \textit{correlation} results for APFD(c) across subjects are given in Table \ref{tab:coupling-correlation}. 

As Figure \ref{fig:coupling-perf} shows, TCP techniques perform differently in terms of detecting different \textit{types} (\eg coupling levels) of real faults. This result yields a few notable observations.  First, TCP techniques tend to perform best (in terms of APFD and APFDc values) on real faults that are classified as needing \textit{stronger operators} to be properly represented by mutants.  This finding is encouraging, as it highlights that the studied approaches are capable of prioritization schemes that effectively uncover faults which are \textit{not} closely coupled to mutants. When examining the correlation results, we find that the APFD $\tau_b$ coefficient values of \textit{coupled} real faults are, unsurprisingly, substantially higher than for other types of real faults. This implies that TCP performance on real faults, which are more tightly coupled to mutants, is more strongly correlated with performance on mutation faults in TCP evaluations. However, for APFDc we find that real faults requiring stronger operators tend to exhibit the highest correlation.  Finally, as Table \ref{tab:coupling-correlation} shows, the correlation results vary across different subject programs. For instance, on one hand, the $\tau_b$ values for Closure are quite large across all levels of coupling, implying that TCP performance on mutants is more indicative of performance on real faults for this particular subject.  On the other hand, the $\tau_b$ values for Lang are much closer to zero. 

\subsubsection{\textbf{Effects of Different Mutation Operators}}

\input{tables/operator-correlation}

\begin{figure}
\setlength{\abovecaptionskip}{-2pt}
\setlength{\belowcaptionskip}{5pt}
\addtolength{\subfigcapskip}{-10pt}
\flushleft
\vspace{-0.5cm}
\subfigure[Closure-2 Bug Fix.\label{fig:example-1}]{
\includegraphics[width=0.46\textwidth]{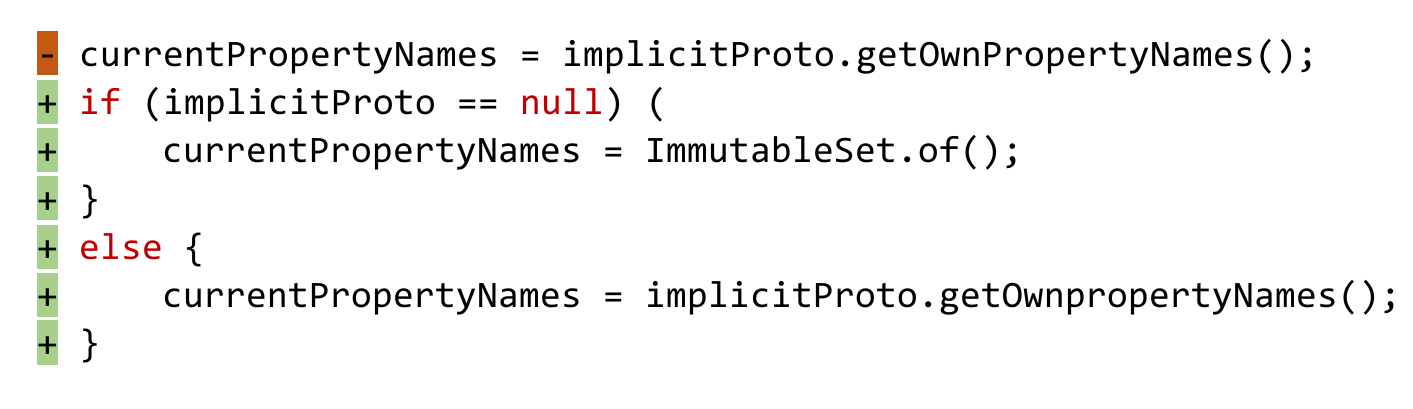}}\\[-0.2cm]
\subfigure[Lang-37 Bug Fix.\label{fig:example-3}]{
\includegraphics[width=0.50\textwidth]{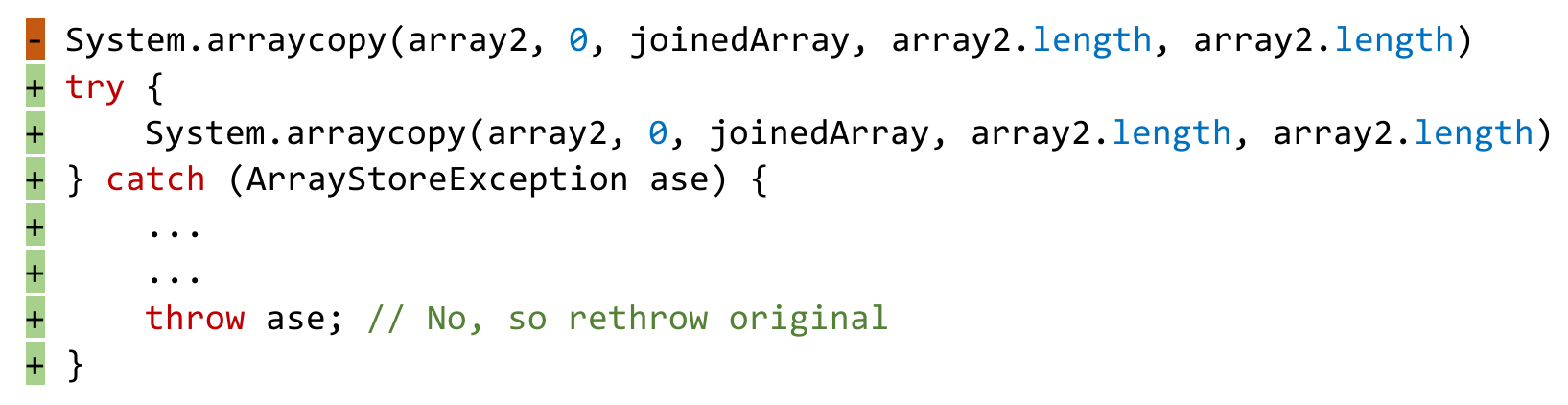}}
\caption{Examples of bug fixing changes.\label{fig:examples}}
\vspace{-0.75cm}
\end{figure}

\begin{figure*}[t]
\vspace{-0.8cm}
\flushleft
\includegraphics[width=0.95\textwidth]{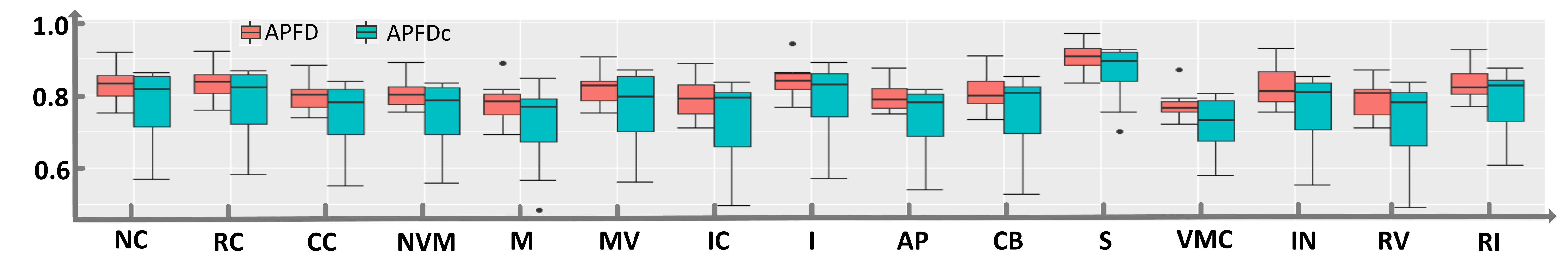}
\vspace{-0.5cm}
\caption{\small Average APFD(c) values across different operators: NC = NegateConditional, RC = RemoveConditional, CC = ConstructorCall, NVM = NonVoidMethodCall, M = Math, MV = MemberVariable, IC = InlineConstant, I = Increments, AP = ArgumentPropagation, CB = ConditionalsBoundary, S = Switch, VMC = VoidMethodCall, IN = InvertNegs, RV = ReturnVals, and RI = RemoveIncrements.}
\label{fig:operator-perf}
\vspace{-0.4cm}
\end{figure*}

The \textit{performance} distributions for the APFD(c) metrics across different operators are depicted as box plots in Figure \ref{fig:operator-perf}.  The \textit{correlation} results across operators between the performance of mutants and real faults are shown in Table \ref{tab:operator-correlation}.  The observations that can be made from this data help further explain the results of RQ$_2$.  The \textit{performance} results illustrate that the different TCP techniques tend to exhibit slight performance variances across different types of mutation operators, with the Switch and VoidMethodCall operators trending toward the positive and negative extremes respectively. This result implies that, even if a researcher is performing \textit{only} a mutation-based analysis, the set of mutation operators selected can cause variations in the results.  More importantly, the \textit{correlation} results indicate that the degree to which performance on mutation faults correlates to performance on real faults, in terms of APFD(c), \textit{varies dramatically} \textit{across} systems as a a whole, and across different operator types \textit{within} a single subject.  For instance, when examining APFD values for specific systems, we found that mutation-based performance for both Closure and Math exhibits a strong positive correlation to performance on real faults across nearly all mutation operators.  At the same time, operators such as \texttt{\small ConstructorCall} and \texttt{\small VoidMethodCall} exhibit a negative correlation within Chart, which tends to have a weak positive correlation overall. Overall the  mutation-based APFDc metric is more strongly coupled to real faults than APFD, corroborating results from RQ$_2$.  
 
Intuitively, our findings suggest that the characteristics of a program may influence how representative mutation-analysis based TCP performance would be in terms of real faults. Therefore, we examined some of the bug fixing commits for a subject program that showed a strong correlation (Closure) and for a program that showed negative correlation (Lang).  When looking into the fixing commits of these two subject programs, we found that Closure, being a compiler, trends heavily toward complex control flows managed by conditionals, compared to Lang, which trends more towards string and array manipulation. Two illustrative examples of bug fixes are shown in Figure \ref{fig:examples}. The first example for Closure (\ie Figure \ref{fig:example-1}) shows that developers fixed this bug by simply modifying conditionals. In particular, the bug shown in Figure \ref{fig:example-1} is exactly the same as one of the PIT mutation operators, \ie the NonVoidMethodCall mutator.  Bug fixing in Lang mostly involved other more complex changes such as adding exception handlers due to the nature of its domain (see Figure \ref{fig:example-3}).  This investigation suggests that TCP performance correlations between real faults and mutants is low when seeded mutants do not properly reflect faults occurring in a given domain or a program.

%% file: tables/overall-correlation.tex
\begin{table}[t]
\small
\centering
\vspace{-0.2cm}
\definecolor{mygray}{gray}{0.9}
\caption{\small Results of the ANOVA analysis and the Kendall $\tau_b$ Coefficient for the overall APFD(c) values shown in Table \ref{tab:overall-perf}.}
\label{tab:overall-correlation}
\vspace{-0.25cm}
\begin{tabular}{|c!{\vrule width 0.8pt}c|c!{\vrule width 0.8pt}c|c|}
\hline
Faults                 & \multicolumn{2}{c!{\vrule width 0.8pt}}{ANOVA $p$-value} & \multicolumn{2}{c|}{$\tau_b$}                         \\ \hline\hline
\multicolumn{1}{|l!{\vrule width 0.8pt}}{} & APFD          & {\cellcolor{mygray}}APFDc        & APFD                   & {\cellcolor{mygray}}APFDc                  \\ \hlinewd{0.7pt}
Real            & 0.011         & {\cellcolor{mygray}}3.22e-4      & - & \cellcolor{mygray} {-}\\ 
Mutant           & 8.02e-4          & {\cellcolor{mygray}}3.77e-5      & 0.143 & \cellcolor{mygray}{0.571} \\ 
S.Mutant           & 0.011     & {\cellcolor{mygray}}1.38e-4      &  -0.071  &    \cellcolor{mygray}0.643             \\\arrayrulecolor{black}\hline
\end{tabular}
\vspace{-0.65cm}
\end{table}

%% file: tables/coupling-correlation.tex
\begin{table*}[t]
\centering
\footnotesize
\vspace{-1.2cm}
\definecolor{mygray}{gray}{0.9}
\caption{Results for the Kendall $\tau_b$ Rank Correlation Coefficient between APFD(c) values for TCP techniques on detecting mutation faults and detecting each type of real faults described in Section \ref{subsubsec:fault-prop}.}
\vspace{-0.3cm}
\label{tab:coupling-correlation}
\begin{tabular}{|c!{\vrule width 0.8pt}c|c!{\vrule width 0.8pt}c|c!{\vrule width 0.8pt}c|c!{\vrule width 0.8pt}c|c!{\vrule width 0.8pt}c|c!{\vrule width 0.8pt}c|c!{\vrule width 0.8pt}}
\hline
Real Faults & \multicolumn{2}{c!{\vrule width 0.8pt}}{Chart}             & \multicolumn{2}{c!{\vrule width 0.8pt}}{Lang}              & \multicolumn{2}{c!{\vrule width 0.8pt}}{Math}              & \multicolumn{2}{c!{\vrule width 0.8pt}}{Time}              & \multicolumn{2}{c!{\vrule width 0.8pt}}{Closure}          & \multicolumn{2}{c!{\vrule width 0.8pt}}{Mean}               \\ \hline\hline
            & APFD   & {\cellcolor{mygray}}APFDc & APFD   & {\cellcolor{mygray}}APFDc & APFD   & {\cellcolor{mygray}}APFDc & APFD   & {\cellcolor{mygray}}APFDc & APFD  & {\cellcolor{mygray}}APFDc & APFD   & {\cellcolor{mygray}}APFDc  \\ \hline
Couple      & 0.429  & {\cellcolor{mygray}}0.786 & -0.071 & {\cellcolor{mygray}}0.286 & 0.5    & {\cellcolor{mygray}}0.429 & -0.143 & {\cellcolor{mygray}}0     & 0.714 & {\cellcolor{mygray}}0.714 & 0.2858 & {\cellcolor{mygray}}0.443  \\ \hline
Limitation  & -0.214 & {\cellcolor{mygray}}0.214 & -0.214 & {\cellcolor{mygray}}0.143 & 0      & {\cellcolor{mygray}}0.286 & -0.071 & {\cellcolor{mygray}}0.143 & 0.5   & {\cellcolor{mygray}}0.286 & 0.0002 & {\cellcolor{mygray}}0.2144 \\ \hline
StrongerOP  & 0.214  & {\cellcolor{mygray}}0.857 & 0.182  & {\cellcolor{mygray}}0.327 & 0      & {\cellcolor{mygray}}0.357 & -0.286 & {\cellcolor{mygray}}0.5   & 0.714 & {\cellcolor{mygray}}0.571 & 0.1648 & {\cellcolor{mygray}}0.5224 \\ \hline
NewOP       & 0.109  & {\cellcolor{mygray}}0.286 & -      & {\cellcolor{mygray}}-     & -0.143 & {\cellcolor{mygray}}0.286 & -      & {\cellcolor{mygray}}-     & 0.571 & {\cellcolor{mygray}}0.571 & 0.179  & {\cellcolor{mygray}}0.381  \\ \hline
\end{tabular}
\vspace{-0.2cm}
\end{table*}

%% file: tables/operator-correlation.tex
\begin{table*}[t]
\centering
\footnotesize
\vspace{-1.0cm}
\definecolor{mygray}{gray}{0.9}
\caption{Results for the Kendall $\tau_b$ Rank Correlation Coefficient between APFD(c) values for TCP techniques on detecting real faults and detecting each type of mutation faults.\vspace{-0.3cm}}
\label{tab:operator-correlation}
\begin{tabular}{|c!{\vrule width 0.8pt}c|c!{\vrule width 0.8pt}c|c!{\vrule width 0.8pt}c|c!{\vrule width 0.8pt}c|c!{\vrule width 0.8pt}c|c!{\vrule width 0.8pt}c|c!{\vrule width 0.8pt}}
\hline
Mutation Faults      & \multicolumn{2}{c!{\vrule width 0.8pt}}{Chart}             & \multicolumn{2}{c!{\vrule width 0.8pt}}{Lang}              & \multicolumn{2}{c!{\vrule width 0.8pt}}{Math}              & \multicolumn{2}{c!{\vrule width 0.8pt}}{Time}              & \multicolumn{2}{c!{\vrule width 0.8pt}}{Closure}          & \multicolumn{2}{c!{\vrule width 0.8pt}}{Mean}               \\ \hline\hline
                     & APFD   & \cellcolor[HTML]{EFEFEF}APFDc & APFD   & \cellcolor[HTML]{EFEFEF}APFDc  & APFD  & \cellcolor[HTML]{EFEFEF}APFDc & APFD   & \cellcolor[HTML]{EFEFEF}APFDc  & APFD  & \cellcolor[HTML]{EFEFEF}APFDc & APFD           & \cellcolor[HTML]{EFEFEF}APFDc          \\ \hline
NegateConditionals   & 0.357  & \cellcolor[HTML]{EFEFEF}0.857 & -0.143 & \cellcolor[HTML]{EFEFEF}0.143  & 0.429 & \cellcolor[HTML]{EFEFEF}0.571 & -0.214 & \cellcolor[HTML]{EFEFEF}0.286  & 0.643 & \cellcolor[HTML]{EFEFEF}0.643 & \textbf{0.214} & \cellcolor[HTML]{EFEFEF}\textbf{0.500} \\ \hline
RemoveConditional    & 0.5    & \cellcolor[HTML]{EFEFEF}0.857 & -0.143 & \cellcolor[HTML]{EFEFEF}0.143  & 0.429 & \cellcolor[HTML]{EFEFEF}0.571 & -0.214 & \cellcolor[HTML]{EFEFEF}0.286  & 0.643 & \cellcolor[HTML]{EFEFEF}0.643 & \textbf{0.243} & \cellcolor[HTML]{EFEFEF}\textbf{0.500} \\ \hline
ConstructorCall      & -0.143 & \cellcolor[HTML]{EFEFEF}0.714 & 0      & \cellcolor[HTML]{EFEFEF}0.286  & 0.5   & \cellcolor[HTML]{EFEFEF}0.714 & -0.214 & \cellcolor[HTML]{EFEFEF}0.071  & 0.714 & \cellcolor[HTML]{EFEFEF}0.5   & \textbf{0.171} & \cellcolor[HTML]{EFEFEF}\textbf{0.457} \\ \hline
NonVoidMethodCall    & 0.214  & \cellcolor[HTML]{EFEFEF}0.786 & 0      & \cellcolor[HTML]{EFEFEF}0.286  & 0.357 & \cellcolor[HTML]{EFEFEF}0.5   & -0.214 & \cellcolor[HTML]{EFEFEF}0.286  & 0.714 & \cellcolor[HTML]{EFEFEF}0.5   & \textbf{0.214} & \cellcolor[HTML]{EFEFEF}\textbf{0.471} \\ \hline
Math                 & 0.286  & \cellcolor[HTML]{EFEFEF}0.714 & -0.286 & \cellcolor[HTML]{EFEFEF}0.286  & 0.286 & \cellcolor[HTML]{EFEFEF}0.5   & -0.071 & \cellcolor[HTML]{EFEFEF}-0.071 & 0.786 & \cellcolor[HTML]{EFEFEF}0.643 & \textbf{0.200} & \cellcolor[HTML]{EFEFEF}\textbf{0.414} \\ \hline
MemberVariable       & 0.214  & \cellcolor[HTML]{EFEFEF}0.929 & -0.786 & \cellcolor[HTML]{EFEFEF}0.357  & 0.429 & \cellcolor[HTML]{EFEFEF}0.5   & 0.286  & \cellcolor[HTML]{EFEFEF}0.357  & 0.5   & \cellcolor[HTML]{EFEFEF}0.357 & \textbf{0.129} & \cellcolor[HTML]{EFEFEF}\textbf{0.500} \\ \hline
InlineConstant       & 0.286  & \cellcolor[HTML]{EFEFEF}0.929 & -0.143 & \cellcolor[HTML]{EFEFEF}0.286  & 0.429 & \cellcolor[HTML]{EFEFEF}0.429 & 0      & \cellcolor[HTML]{EFEFEF}0.071  & 0.786 & \cellcolor[HTML]{EFEFEF}0.571 & \textbf{0.272} & \cellcolor[HTML]{EFEFEF}\textbf{0.456} \\ \hline
Increments           & 0.214  & \cellcolor[HTML]{EFEFEF}0.714 & -0.214 & \cellcolor[HTML]{EFEFEF}0.143  & 0.286 & \cellcolor[HTML]{EFEFEF}0.571 & 0      & \cellcolor[HTML]{EFEFEF}0      & 0.786 & \cellcolor[HTML]{EFEFEF}0.714 & \textbf{0.214} & \cellcolor[HTML]{EFEFEF}\textbf{0.428} \\ \hline
ArgumentPropagation  & 0.143  & \cellcolor[HTML]{EFEFEF}0.857 & 0      & \cellcolor[HTML]{EFEFEF}0.214  & 0.357 & \cellcolor[HTML]{EFEFEF}0.286 & -0.429 & \cellcolor[HTML]{EFEFEF}0.286  & 0.643 & \cellcolor[HTML]{EFEFEF}0.643 & \textbf{0.143} & \cellcolor[HTML]{EFEFEF}\textbf{0.457} \\ \hline
ConditionalsBoundary & 0.357  & \cellcolor[HTML]{EFEFEF}0.786 & -0.071 & \cellcolor[HTML]{EFEFEF}0.286  & 0.429 & \cellcolor[HTML]{EFEFEF}0.643 & 0.071  & \cellcolor[HTML]{EFEFEF}0.214  & 0.714 & \cellcolor[HTML]{EFEFEF}0.571 & \textbf{0.300}   & \cellcolor[HTML]{EFEFEF}\textbf{0.500}   \\ \hline
Switch               & 0.214  & \cellcolor[HTML]{EFEFEF}0.714 & -0.214 & \cellcolor[HTML]{EFEFEF}-0.071 & 0.429 & \cellcolor[HTML]{EFEFEF}0.357 & -0.214 & \cellcolor[HTML]{EFEFEF}0.214  & 0.786 & \cellcolor[HTML]{EFEFEF}0.429 & \textbf{0.200} & \cellcolor[HTML]{EFEFEF}\textbf{0.329} \\ \hline
VoidMethodCall       & -0.143 & \cellcolor[HTML]{EFEFEF}0.714 & -0.214 & \cellcolor[HTML]{EFEFEF}0.071  & 0.214 & \cellcolor[HTML]{EFEFEF}0.571 & -0.357 & \cellcolor[HTML]{EFEFEF}0.357  & 0.857 & \cellcolor[HTML]{EFEFEF}0.643 & \textbf{0.071} & \cellcolor[HTML]{EFEFEF}\textbf{0.471} \\ \hline
InvertNegs           & 0.357  & \cellcolor[HTML]{EFEFEF}0.857 & 0.143  & \cellcolor[HTML]{EFEFEF}0.071  & 0.143 & \cellcolor[HTML]{EFEFEF}0.5   & -0.214 & \cellcolor[HTML]{EFEFEF}0      & 0.714 & \cellcolor[HTML]{EFEFEF}0.714 & \textbf{0.229} & \cellcolor[HTML]{EFEFEF}\textbf{0.428} \\ \hline
ReturnVals           & 0.357  & \cellcolor[HTML]{EFEFEF}0.857 & -0.429 & \cellcolor[HTML]{EFEFEF}0.357  & 0.357 & \cellcolor[HTML]{EFEFEF}0.714 & -0.214 & \cellcolor[HTML]{EFEFEF}0.286  & 0.786 & \cellcolor[HTML]{EFEFEF}0.571 & \textbf{0.171} & \cellcolor[HTML]{EFEFEF}\textbf{0.557} \\ \hline
RemoveIncrements     & 0.214  & \cellcolor[HTML]{EFEFEF}0.786 & -0.143 & \cellcolor[HTML]{EFEFEF}0.071  & 0.429 & \cellcolor[HTML]{EFEFEF}0.429 & 0.143  & \cellcolor[HTML]{EFEFEF}0.071  & 0.714 & \cellcolor[HTML]{EFEFEF}0.714 & \textbf{0.271} & \cellcolor[HTML]{EFEFEF}\textbf{0.414} \\ \hline
\end{tabular}
\vspace{-0.5cm}
\end{table*}

%% file: threats.tex

\section{Threats to Validity}
\label{sec:threats}

\noindent
\textbf{Threats to Internal Validity} concern potential confounding factors of the experiments that might introduce observed effects. One such factor is represented by faults that were seeded into the programs.  We chose PIT to perform mutation analysis, which contains different operators compared to other mutation  tools, such as Major \cite{Major}. 
While PIT features many operators common across other tools, it	 is possible that different mutation tools might have led to different observations. We leave exploration of additional tools as future work.

	Internal validity threats also arise due to the assumptions made about the validity of the coupling between mutants and real faults obtained from Just \etal \cite{Just:FSE14}. This is because mutants seeded in the same code where the fault occurred may differ from the real fault, hence leading to possible mis-classifications. Future research could further examine the affects of such coupling relationships to mitigate this threat. 

Another potential confounding factor is the fact the studied test suites are written by developers.  However, these test suites have been shown by prior work \cite{Just:FSE14} to generally be of high quality, exhibiting high coverage, mitigating this threat.  Additionally, threats may arise due to the utilization of the same test suite between program versions containing real faults and the latest program version (to which mutants were seeded).  
However, previous studies illustrate that the performance of mutation-based TCP techniques tend to be similar across different program versions for the same test suite~\cite{Lu:ICSE16,Henard:ICSE16}.

\noindent
\textbf{Threats to Construct Validity}  concern the relation between experimental theory/constructs, and potential effects on observed results.  As explained in Section \ref{subsec:methodology}, in the context of this study we re-implemented all of the TCP techniques following the approach descriptions in their respective papers.  Our re-implementations may differ slightly from the original versions. However, we closely followed the methodology of the previous work, and three authors and one external TCP expert reviewed the code to ensure a reliable implementation. Executing the studied TCP techniques on all of the program versions was time-consuming, totaling more than five months of computation time. To make the GA-based technique tractable we reduced the maximum number of generations to 50. Also, to allow for a fair comparison, we used the same GA settings for both real faults and mutants.

\noindent
\textbf{Threats to External Validity.} We limited our focus to eight TCP techniques, which require only source code, test code, and coverage information to perform prioritization. These eight TCP techniques are well-understood and widely used/studied in recent research work \cite{Luo:FSE16,Lu:ICSE16}, and since we aimed to understand how techniques differed from previous studies when applied to real faults, this set of techniques is suitable. We encourage researchers to extend this study to additional TCP approaches \revision{and technique configurations.}

	In order to provide a rigorous experimental procedure, we applied mutation analysis to TCP techniques in particular experimental settings discussed in Sec. \ref{subsec:context} and \ref{subsec:methodology}.  Thus, it is possible that these results may differ for other TCP evaluation methodologies. However, we chose the experimental methodology set forth in this paper due to the fact that it has been widely used in previous studies \cite{Hao:TOSEM14,Ledru:ASE12,Lu:ICSE16,Luo:FSE16,Mei:TSE12,Zhang:ASE13} and is likely to be used in the future. 

We use the Defects4J dataset to understand the effectiveness of TCP techniques in terms of real-fault detection. It is possible that there are different types of faults (varying in complexity) in other subject programs written in other program languages compared to those in Defects4J. However, Defect4J is one of the largest and most studied~\cite{Just:ISSTA14,Just:FSE14,Shamshiri:ASE15} publicly available databases of real faults, containing 357 faults extracted from real-world software systems.  

	We utilize Pit \cite{PIT} and hence our results are representative of a certain set of mutants.  While Pit utilizes many of the same standard operators as other tools, this study could be expanded in the future to consider additional mutation testing tools.

%% file: lessons.tex
\section{Lessons Learned} 
\label{sec:lessons}

In this section we summarize the pertinent findings of our study into discrete \textit{learned lessons} and discuss their potential impact on future work in the TCP area. 

\noindent
\textbf{Lesson 1:} \textit{Relative Performance of TCP techniques on mutants may not indicate similar relative performance on real faults, depending on the subject program.}  Our study indicates that, for the subject programs studied, the relative performance of TCP techniques (following the popular methodology utilized in our experiments) is not similar between mutants and real faults. This indicates that a technique which outperformed competing techniques under the experimental setting of mutation analysis may not outperform competing techniques on real faults. This highlights a potential threat to validity for mutation-based assessments of TCP approaches, impacting the generalizability of TCP performance comparisons to real program faults.  Future work should proceed in two directions. First, techniques for carefully selecting mutants should be pursued (see Lesson 3). Second, there is a clear need for comprehensive datasets of real faults, such as Defects4J, in order to more thoroughly evaluate TCP approaches.  Therefore, researchers should focus on developing reliable automated or semi-automated techniques to extract and isolate real faults from existing open source software projects, which clearly calls for a community-wide effort.

\noindent
\textbf{Lesson 2:} \textit{The metrics utilized in mutation-based evaluations of TCP techniques impact the representativeness of performance on real faults.}  We found that mutation-based APFD values generally exhibit only a \textit{weak positive correlation} to APFD values calculated in terms of real fault discovery, whereas for APFDc this correlation was medium to strong.  However, such results varied across subjects programs.  This is important as it signals that when considering the extremely popular APFD metric, mutation-based performance of a particular TCP technique \textit{may} be independent of its performance on real faults.  This means that, depending on the subject program, one may not be able to rely upon mutation-based APFD to predict the \textit{practical} performance of TCP technique on real faults.  While one could use the more strongly correlated APFDc metric, researchers may not always want to include execution cost in their performance evaluation.

\noindent
\textbf{Lesson 3:} \textit{The types of mutation operators utilized for TCP performance evaluations must be carefully selected or derived in order for the results to be representative of performance on real faults.}  Our results indicate that correlations in TCP performance between mutants and real faults vary both across subject programs and across different types of mutation operators within a specific subject program. This suggests that different characteristics of subject programs most likely play a role in determining the representativeness of  certain mutation operators for a particular subject (or domain).  This is actually a positive outcome when considering the future applicability of mutation testing to TCP evaluations, as it shows that \textit{under the right circumstances} mutation-based TCP can, in fact, be realistic.  However, in order to properly achieve the ``correct circumstances" for mutation operators to be applied, further research needs to be conducted.  This specifically illustrates the need for the following interconnected research threads: (i) deriving, either manually or automatically, fault models for specific software systems or domains; (ii) developing tailored mutation operators based on such fault models; and (iii) seeding mutants using rigorous statistical methods according to observed distributions of faults.  If thoroughly pursued, we believe that research efforts directed toward these goals will provide for future tools capable of generating mutants that are more representative of real faults, not only in for TCP, but also in other areas of software testing.

%% file: conclusion.tex
\vspace{-0.1cm}
\section{Conclusion} 
\label{sec:concl}
\vspace{-0.1cm}

In this work we conducted the first empirical study investigating the extent to which mutation-based evaluations of TCP approaches are \textit{realistic}.  We examined the performance, in terms of both the APFD and APFDc metrics, of eight different TCP approaches applied to a dataset of 357 real world faults from the Defects4J dataset and a set of over 35k mutants.  Our results indicate that typical mutation-based evaluations of TCP techniques tend to \textit{overestimate} absolute performance on real faults. Furthermore, for the APFD metric, relative performance on mutants may not be representative of relative performance on real faults, depending upon mutant and program characteristics.  These findings highlight the need for future work on deriving mutation operators tailored toward specific subject programs or domains, driving mutation-based TCP evaluations toward being more realistic.